\documentclass[12pt]{iopart}
\usepackage{graphicx}
\usepackage{dcolumn}
\usepackage{bm}

\begin{document}

\title{Parametrically excited ``Scars'' in Bose-Einstein condensates}

\author{Nadav Katz and Oded Agam}
\address{The Racah Institute of Physics, The Hebrew University, Jerusalem, 91904, Israel}
\ead{agam@phys.huji.ac.il}

\date{\today}

\begin{abstract}
Parametric excitation of a Bose-Einstein condensate (BEC) can be realized by periodically changing the
interaction strength between the atoms. Above some threshold strength, this excitation modulates the condensate density. We show that when the condensate is trapped in a potential well of irregular shape, density waves can be strongly concentrated ("scarred") along the shortest periodic orbits of a classical particle moving within the confining potential. While single-particle wave functions of systems whose classical counterpart is chaotic may exhibit rich scarring patterns, in BEC, we show that nonlinear effects select mainly those scars that are locally described by stripes. Typically, these are the scars associated with self retracing periodic orbits that do not cross themselves in real space. Dephasing enhances this behavior by reducing the nonlocal effect of interference.
\end{abstract}

\maketitle

\section{Introduction}

One of the hallmarks of quantum systems whose classical counterpart exhibits chaotic motion is the "scar" phenomenon, i.e. the concentration of wave functions along the short periodic orbits of the underlying classical dynamics \cite{Heller}. Scars commonly appear in wave equations of quantum chaotic systems such as a single particle moving within a potential well whose boundary is in the form of a stadium, and otherwise flat. The  classical counterpart of this system is a particle specularly reflected from the boundary. Therefore this type of single-particle two-dimensional systems, have been called "billiards". Most studies of scars focus on billiards. However scars may also appear in many-body systems, where additional factors come into play such as the interparticle interactions and dephasing effects.

In this work we analyze scars in the density excitations of Bose-Einstein condensates (BEC) of ultracold atoms \cite{RaizenChaos}, confined to a potential well of irregular shape similar to a billiard (see e.g. Fig. 1), and parametrically excited by a periodic change of the interaction strength between the atoms. Recently it has been shown, both theoretically \cite{FaradayTheory,Solitons} and experimentally \cite{FaradayExperiment}, that such an excitation of BECs generates waves similar to surface Faraday waves in a periodically accelerated fluid. With advances in trapping and imaging technology, bulk parametric excitations are now feasible and the density variations they generate may be observed in-situ \cite{in-situ}.
\begin{figure}[t]
\centering
\includegraphics[width=8.5cm]{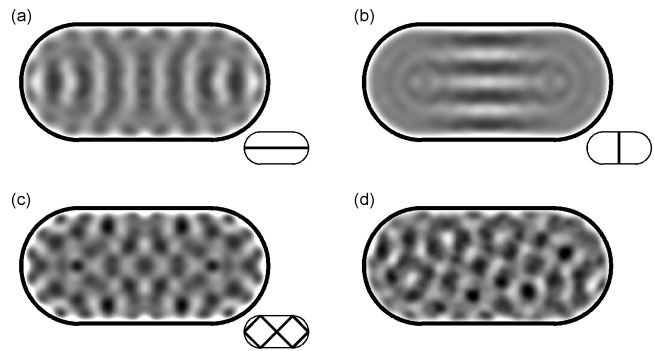}
\caption{Steady-state density profiles of parametrically driven condensates trapped in a stadium billiard, as obtained from a numerical solution of Eq.~(\ref{eq:GP}) at various excitation frequencies \cite{Parameters}.
(a) The "horizontal" scar. (b) The "bouncing ball" scar. (c) The "bow-tie" scar.  At the lower right side of these panels we depict the corresponding periodic orbit. Panel (d) represents an example of the density profile of the BEC in the presence of dephasing. The parameters of this figure are identical to those of panel (c) but with short range gaussian correlated noise added to the confining potential. }
\label{Fig1}
\end{figure}

This system can be realized, for example, by using a rapidly moving laser beam or a hologram which generates a potential well \cite{DavidsonTraps}, $V({\bf r})$. This potential confines the condensate to a cylinder whose cross-section, in the transverse direction ${\bf r}= (x,y)$, may be shaped as a chaotic billiard.  Additional laser beams are used to confine the condensate in the perpendicular direction, ${\bf z}$, in order to form a quasi-two dimensional geometry \cite{Petrov2D,Ketterle2D}. The parametric excitation of the condensate is obtained either by modulating the confining potential in the ${\bf z}$ direction\cite{Juan1999,Salasnich02}, or by applying a time dependent magnetic field near a Feshbach resonance \cite{KetterleFeshbach,Kevrekidis03}. Both effectively introduce a time dependent component into the interaction energy between the atoms.

Assuming that the excitations involve a large number of atoms, the system may be modeled by the Gross-Pitaevskii equation,
\begin{eqnarray} i \hbar\frac{ \partial \psi}{\partial t}&=&(1\!-\! i \Gamma) \left( -\frac{\hbar^2}{2 m}\nabla^2 \!+\! V({\bf r})\!-\! \mu + \bar{g} |\psi|^2 \right) \psi + \delta g(t)|\psi|^2 \psi,
\label{eq:GP}
\end{eqnarray}
where $\Gamma$ is a dimensionless phenomenological parameter accounting for the dissipation \cite{pitaevskii58}, $m$ is the effective mass of the atoms, $\nabla^2= \partial_x^2+\partial_y^2$ is the two-dimensional Laplacian, $\mu$ is the chemical potential,
$\bar{g}=\frac{4\pi a \hbar^2}{m d_z}$ denotes the bare contact interaction strength, where $a$ is the $s$-wave scattering length and $d_z$ is the effective width of the confining potential in the  transverse ($z$) direction\cite{Salasnich02}. $\delta g(t)= 2 a \cos(2\omega t)$ represents a small time dependent modulation of the contact interaction, with frequency $2\omega$ and a small amplitude $2a \ll \bar{g}$.

In Fig.~1 we depict a few examples of the excited condensate density, $|\psi({\bf r},t)|^2$ at long times where a steady state excitation is reached. These results are obtained by numerical solution of Eq.~(\ref{eq:GP}) in a stadium billiard potential\cite{Parameters}. Our numerical study indicates that the shortest scars,  shown in panels (a) and (b), are commonly observed, while the "bow-tie" scar associated with a self crossing orbit, shown in panel (c), rarely appears, as it requires a fine tuning of the parameters. We have not been able to observe parametric excitation of other scars or random wave functions. This behavior, as we show below, results from interaction effects associated with the nonlinear term in the Gross-Pitaevskii equation (\ref{eq:GP}).

This is the plan of the paper: In section \ref{sec:Linear_analysis} we present the stability analysis of the problem, and show that it is described by the dissipative Mathieu's equation. Next, in Section \ref{sec:nonlinear_analysis}, we focus our attention on the leading instability of the dissipative Mathieu equation, and identify the most relevant nonlinear terms of the equations of motion. We solve these equations in order to find a steady state solution of the problem describing an excitation of a single mode. In section \ref{sec:scars} we consider the case where the strength of excitation is sufficiently large to excite several modes, and show that interactions induce mode locking which promote the appearance of short scars with local density pattern in the form of stripes. The effect of noise and dephasing on the excitation patterns will be discussed in section \ref{sec:noise_and_dephasing}. Finally, in section \ref{sec:numerical_study} we present a numerical study of the problem, and in \ref{sec:summery} we conclude.

\section{Linear analysis}
\label{sec:Linear_analysis}

Our analysis will be limited to the experimentally relevant regime where the Thomas-Fermi approximation is valid.  Namely when the healing length, $\xi=\hbar/\sqrt{m \mu}$, is much smaller than the size of the system, $L$. Under this condition, the condensate density, $|\psi({\bf r},t)|^2$, in the absence of pumping, i.e. $\delta g=0$, is essentially constant within the billiard, and equals to $|\bar{\psi}|^{2}= \mu/\bar{g}$.

When the parametric pumping is applied weakly enough, it only modifies the phase of the wave function. Thus $\psi(t)= |\bar{\psi}| e^{i\bar{\phi}(t)}$ where $\bar{\phi}(t) = \frac{\mu}{\hbar \bar{g}} \int^t dt' \delta g(t')$. However, as the strength of the excitation increases, the uniform density solution becomes unstable, and a parametric resonance sets in. In order to describe the behavior of the system in this regime, we follow the standard Bogoliubov analysis \cite{Bogoliubov} in the presence of the time dependent drive and dissipation.

Let us assume the excitation strength to be sufficiently close to the instability threshold, and expand the wave function in the form,
\begin{eqnarray} \psi({\bf r},t)= |\bar{\psi}|e^{i \bar{\phi}(t)}  \left[ 1+ \sum_j w_j(t) u_j({\bf r})\right], \label{eq:ansatz}
\end{eqnarray}
where $w_j(t)$ are complex amplitudes, while $u_j({\bf r})$ are the eigenstates of the "unperturbed Hamiltonian",
\begin{eqnarray}
\left[ -\frac{\hbar^2}{2m}\nabla^2 + V({\bf r}) \right] u_j({\bf r})= \epsilon_j u_j({\bf r}), \label{eq:eigenstates}
\end{eqnarray}
normalized such that $\int d^2 r |u_j({\bf r})|^2=1$. From here on we choose these eigenstates to be real functions.

Substituting Eq.~(\ref{eq:ansatz}) in (\ref{eq:GP}), multiplying the result by $u_j({\bf r})$, and integrating over space, we obtain a set of coupled nonlinear equations for the expansion coefficients, $w_j(t)$. Yet in its linearized form, this set of equations decouples, such that
\begin{eqnarray}
\frac{\partial}{\partial t} \left( \begin{array}{c} w_j \\ w_j^* \end{array} \right) = M \left( \begin{array}{c} w_j \\ w_j^* \end{array} \right)+ O(w_{j'}^2), \label{eq:linear}
\end{eqnarray}
where $*$ denotes complex conjugation, and
\begin{eqnarray}
M=-\frac{1}{\hbar}\left[ \begin{array}{cc}
(\Gamma\!+\!i)(\epsilon_j\!+\!\mu)\!+\! i\frac{\mu \delta g}{\bar{g}}  & (\Gamma+i)\mu+i\frac{\mu \delta g }{\bar{g}} \\ (\Gamma-i)\mu-i\frac{\mu
\delta g}{\bar{g}}  & (\Gamma\!+\!i)(\epsilon_j\!+\!\mu)\!+\!i\frac{\mu \delta g}{\bar{g}}  \end{array} \right].
\end{eqnarray}

The above equations reduce to the damped Mathieu equation,
\begin{eqnarray}
\frac{d^2\eta_j}{dt^2}+2 \gamma_j \frac{d\eta_j}{dt}+\left[ \Omega_j^2+ 4 h_j \cos (2\omega t) \right]\eta_j=0, \label{eq:Mathieu}
\end{eqnarray}
where $\eta_j= \mbox{Re}~ w_j$ is the real part of the amplitude of the  j-th mode, $\gamma_j= \Gamma(\mu+\epsilon_j)/\hbar$ is its effective damping rate, $\Omega_j^2= (\epsilon_j(\epsilon_j+2 \mu)(1+\Gamma^2))/\hbar^2$, is the frequency, and $h_j= a \mu \epsilon_j/(\hbar^2 \bar{g})$ is the effective pumping amplitude. The stability plane of Mathieu's equation is depicted in Fig.~2. The vertical axis represents the excitation strength while the horizontal axis is the ratio of the natural frequency of the oscillator, $\Omega_j$ to half of the pumping frequency, $\omega$. The dark tongues on this plot show the regions where the system is unstable, namely the values of parameters where the amplitude, $w_j$, grows exponentially in time.  It is evident that the leading instability, namely the first one encountered as the excitation increases, is associated with the tongue where half of the pumping frequency is close to one the frequencies, $\Omega_j$.

\begin{figure}[t]
\centering
\includegraphics[width=8.0cm]{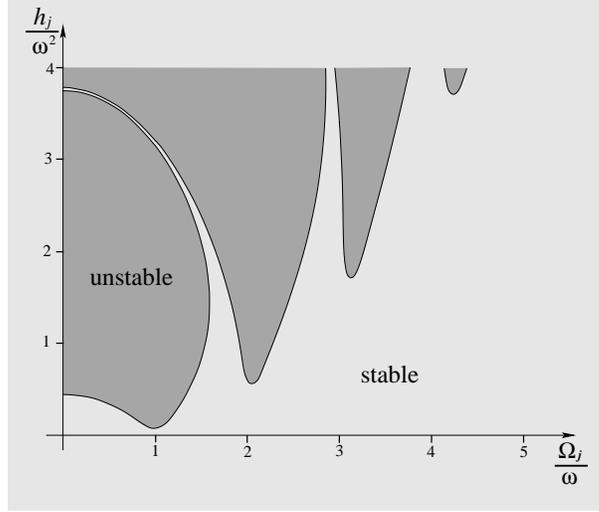}
\caption{A schematic illustration of the stability plane of the damped Mathieu equation (\ref{eq:Mathieu}) at small damping coefficient $\Gamma \ll 1$.}
\label{Fig2}
\end{figure}

>From here on we shall therefore assume that the pumping strength is sufficiently weak and focus on the main instability where $\omega \sim \Omega_j$.
In order to describe the behavior of the system in this case, let us perform a Bogoliubov-like \cite{Bogoliubov} pseudo-rotation which diagonalizes the above matrix for the case of zero pumping strength,
\begin{eqnarray}
\left( \begin{array}{c} b_j \\ b_j^* \end{array} \right) = U^{-1} \left( \begin{array}{c} w_j \\ w_j^* \end{array} \right),
\end{eqnarray}
where
\begin{eqnarray} U= \left( \begin{array}{cc} \cosh \left( \frac{\theta+ i \theta'}{2}\right) & -\sinh \left( \frac{\theta-i\theta'}{2}\right) \\ -\sinh \left(\frac{\theta+ i\theta'}{2}\right) &
\cos \left( \frac{\theta- i\theta'}{2} \right) \end{array} \right),
\end{eqnarray}
with
\begin{eqnarray} \theta =  \tanh^{-1}\left( \frac{\mu}{\mu+\epsilon_j} \right), ~~ \theta ' =  \sin^{-1}\left( \frac{\mu\Gamma}{\epsilon_j(\epsilon_j+2\mu)} \right).
\end{eqnarray}
Thus, the vector $(b_j,b_j^*)$ satisfies the equation
\begin{eqnarray}
\frac{\partial}{\partial t} \left( \begin{array}{c} b_j \\ b_j^* \end{array} \right) = U^{-1} M U \left( \begin{array}{c} b_j \\ b_j^* \end{array} \right)+ O(w_{j'}^2), \label{eq:linear-b}
\end{eqnarray}
Assuming the pumping  strength to be sufficiently close to the excitation threshold, one may write the amplitudes in the form $b_j(t)=B_j(t) e^{-i\omega t}$, where $B_j(t)$ is a slowly varying function with respect to the frequency $\omega$. Averaging the equations (\ref{eq:linear-b}) over the rapid time oscillations leads to:
\begin{eqnarray}
\frac{\partial}{\partial t} \left( \begin{array}{c} B_j \\ B_j^* \end{array} \right) = M_B \left( \begin{array}{c} B_j \\ B_j^*
\end{array} \right)+  O(B_{j'}^2), \label{eq:linearB}
\end{eqnarray}
where
\begin{eqnarray}
M_B=\left[ \begin{array}{cc} (-\gamma_j+i(\omega-\omega_j)
& -i h_j e^{i\theta'} \\ i h_j e^{- i\theta'}  & (-\gamma_j-i(\omega-\omega_j)  \end{array} \right], \label{eq:MB}
\end{eqnarray}
and
\begin{eqnarray}
\omega_j^2= \Omega_j^2- \gamma_j^2= \left[\epsilon_j(\epsilon_j+2 \mu)-\Gamma^2\mu^2 \right]/\hbar^2.
\end{eqnarray}

Seeking a solution where both $B_j$ and $B_j^*$ behave like $e^{\sigma_j t}$, one obtains that the growth exponent, $\sigma_j$, is given by
\begin{eqnarray}
 \sigma_j^{(\pm)}=-\gamma_j \pm \sqrt{ h_j^2-(\omega-\omega_j)^2}.
 \end{eqnarray}
 Thus the uniform density solution becomes unstable when $\sigma_j^{(+)} >0$, i.e.
 \begin{eqnarray}
 h_j> \sqrt{\gamma_j^2+(\omega-\omega_j)^2}. \label{eq:EXthreshold}
 \end{eqnarray}
 In particular, when half of the excitation frequency coincides with one of the  eigenfrequencies of the system, $\omega=\omega_j$, the criteria for the onset of instability is  $h_j= \gamma_j$, or alternatively in terms of the excitation amplitude $a=\hbar \bar{g} \Gamma(\mu+\epsilon_j)\omega_j/(\mu \epsilon_j)$.

\section{Nonlinear analysis}
\label{sec:nonlinear_analysis}
Above threshold, the system oscillates at half of the pumping frequency, and its amplitude grows exponentially with time. Dissipation does not saturate the growth of the amplitude (it only shifts the excitation threshold), and one must resort to the nonlinear terms of the governing equations in order to describe the evolution over long time scales.

To identify the most important nonlinear terms of Eq.~(\ref{eq:linear}), notice that these may be divided into two categories: Quadratic terms, $w_{j'}(t) w_{j''}(t)$, proportional to integrals of the form $\int d^2 r u_j({\bf r}) u_{j'}({\bf r})u_{j''}({\bf r})$, and cubic terms,
$w_{j'}(t) w_{j''}(t)w_{j'''}(t)$, proportional to $\int d^2 r u_j({\bf r}) u_{j'}({\bf r})u_{j''}({\bf r})u_{j'''}({\bf r})$. At high excitation frequencies, the spatial oscillatory nature of the eigenstates  $u_{j}({\bf r})$ implies that for any choice of the indices $j$, $j'$, and $j''$,
the quadratic contribution is
small compared to the cubic terms where either one of the following conditions is satisfied: (a) $j=j'$ and $j''=j'''$; (b) $j=j''$ and $j'=j'''$; (c) $j=j'''$ and $j'=j''$. Other cubic terms will also be small. 

Therefore, from here on, we shall keep only the leading nonlinear terms, an approach which is similar in spirit to the Hartree-Fock approximation. If furthermore we neglect the nonlinear terms that are proportional to $\delta g(t)$, and express the result in terms of the slowly changing amplitudes, $B_j$, we obtain that
\begin{eqnarray}
\partial_t B_j= (\partial_t B_j)_{\mbox{lin}}+(\partial_t B_j)_{\mbox{int}}. \label{eq:B_nonlinearEq}
\end{eqnarray}
Here
$(\partial_t B_j)_{\mbox{lin}}$ represents the linear contribution expressed by Eqs.~(\ref{eq:linearB}) and (\ref{eq:MB}), while
\begin{eqnarray}
\left(\frac{\partial B_j}{\partial t}\right)_{\mbox{int}}&=&-\frac{(i+\Gamma)\mu}{\hbar} \sum_{j'} I_{jj'}\frac{\mu+\epsilon_j }{\sqrt{\epsilon_j(\epsilon_j+2\mu)}} \nonumber\\ & \times &\left[ 2(1-\delta_{j,j'})|B_{j'}|^2B_j + B_{j'}^2 B_j^*\right] \label{eq:int}, \end{eqnarray}
 where
 \begin{eqnarray} I_{jj'}= \int d^2r u_j^2({\bf r})u_{j'}^2({\bf r}).  \label{eq:InteractionTerms}
 \end{eqnarray}
is the nonlinear contribution which couples $B_j$ to other modes, $B_{j'}$

Let us consider first the steady state solution, $\partial_t B_j=0$, where only one mode, say the $j$-th one, is excited. Equation (\ref{eq:B_nonlinearEq}) in this case reduces to:
\begin{eqnarray} \left[-\gamma_j + i(\omega-\omega_j)\right] B_j -i h_je^{i\theta '} B_j^* -
 \frac{(i+\Gamma)\mu(\mu+\epsilon_j)}{\hbar\sqrt{\epsilon_j(\epsilon_j+2\mu)}} I_{jj} |B_j|^2 B_j=0,  \label{eq:SS}
 \end{eqnarray}
and its solution for $N_j= |B_j|^2$ is
\begin{eqnarray} N_j = \frac{\hbar}{I_{jj}} \frac{ \sqrt{\epsilon_j(\epsilon_j+2
\mu)}}{\mu(\mu+\epsilon_j)(1+ \Gamma^2)}~~~~~~~~~~~~~~~~~~~~~~~~~~~~~~~~~~~ \label{eq:main} \\ \times
 \left[\sqrt{ h_j^2(1+\Gamma^2)-(\gamma_j+\Gamma\omega- \Gamma \omega_j)^2} -\Gamma \gamma_j +\omega -\omega_j
\right]\nonumber.
 \end{eqnarray}
This solution is valid above the excitation threshold (\ref{eq:EXthreshold}), otherwise $N_j=0$.

 $N_j$ is roughly proportional to the number of atoms in the excited mode $u_j({\bf r})$. Its relation to the spatial variation of the BEC density follows from
 \begin{eqnarray}
 \delta \rho({\bf r},t)&=& |\bar{\psi}(1+ w_j(t) u_j({\bf r})|^2- |\bar{\psi}|^2  \nonumber \\
 &\simeq & 2|\bar{\psi}|^2 u_{j}({\bf r})~\mbox{Re} ~w_j(t) ,
 \end{eqnarray}
 where
\begin{eqnarray}
  \mbox{Re}~ w_j(t)= \frac{ \sqrt{N_j}}{(1+ 2\mu/\epsilon_j)^{1/4}} \cos (\omega t- \phi_j).
 \end{eqnarray}
Here $\phi_j$ is the phase shift between the pump and the excited wave which may be computed from the solution of Eq.~(\ref{eq:SS}). In particular, in the limit $\Gamma \to 0$ it approaches $\pi/2$, implying that the saturation mechanism is that of phase mismatch between the pump and the system.

\section{Scars}
\label{sec:scars}

Until now we have considered only the case where one mode is excited. However, usually several modes will be simultaneously excited when the system is large enough. The reason is that the frequency difference between neighboring modes (being inversely proportional to the system's area)  is  small compared to the width of the leading instability tongue near the threshold, and therefore near the minimum of the instability tongue (see Fig.~3) the condition (\ref{eq:EXthreshold}) will usually be satisfied for several eigenfrequencies, $\omega_j$.

For sufficiently weak excitation and a fine tuning of the frequency, it is possible to excite a single eigenstate of the system. However, in this case, the growth rate exponent is extremely small, and it takes rather long time for the system to reach a steady state. During this time, external noise (discussed later on) and  non-parametric excitations, generated near the edge of the system, will overwhelm the response.

We shall therefore consider the more practical situation where several modes are simultaneously excited. Our numerical simulations, as well as experimental data for Faraday waves in hydrodynamic systems \cite{Kudrolli}, indicate that in this case the natural density patterns that are parametrically excited are the short scars of the system.  Scars are not exact eigenstates of the system. However they may be viewed as resonances whose width is determined by the typical time that the underlying classical particle stays in the vicinity of the periodic orbit associated with the scar \cite{Bogomolny1988}.

The set of frequencies, $\omega_\nu^{(l)}$, associated with a given scar (denoted by the subscript $\nu$), approximately satisfies the quantization condition of a particle in a one dimensional square well potential whose width equals to the length of the scar, namely, $\omega_\nu^{(l)} \simeq \frac{c}{L_\nu} (2\pi l-\phi_\nu)$ where $c$ is the sound velocity, $L_\nu$ is the total distance of a single round trip along the periodic orbit, $l$ is an integer, and $\phi_\nu$ is the the Maslov phase accumulated along the periodic orbit (e.g. due to bouncing from the boundary of the well and passing through focusing points). Thus the frequency gap between two neighboring frequencies of short scars, $\Delta \omega_\nu= \omega_\nu^{(l)}-  \omega_\nu^{(l-1)}= 2\pi c/L$  is, in general, much larger than the mean frequency spacing of the exact eigenmodes of the system, as illustrated in Fig.~3.

\begin{figure}[t]
\centering
\includegraphics[width=8.0cm]{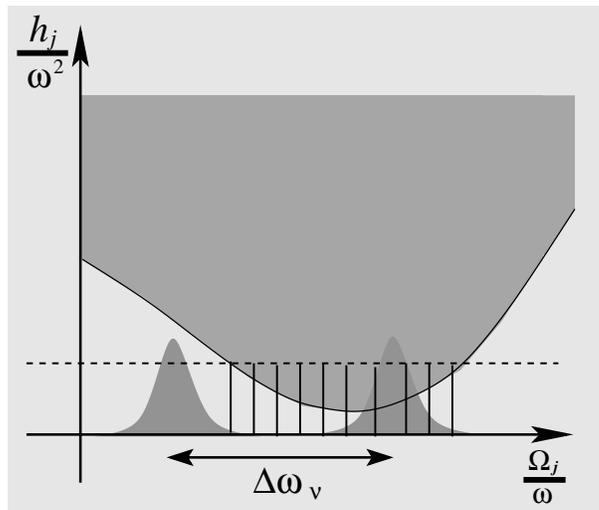}
\caption{A magnification of the leading instability tongue shown in Fig.~2, and the eigenfrequencies of the systems (designated by the thin black bars. When the excitation strength $a$  is sufficiently high, a large number of modes are excited. The gray Lorenzian-like curves represent the resonances associated with some scar.}
\label{Fig3}
\end{figure}

In order to understand how scars may emerge from the simultaneous excitation of several states, it is instructive to describe the general structure of the wave functions of chaotic systems. These may be roughly viewed as a combination of two components: The first, $\psi_{rand}$ is approximately random Gaussian wave function which results from a superposition of plane waves in arbitrary directions and arbitrary phases\cite{berry77}. This contribution  may be associated with long orbits of the underlying classical dynamics and is characterized by universal statistical properties. The second component, $\psi_{scar}$ is nonuniversal since it emerges from constructive interference along the shortest periodic orbits of the classical system. This contribution leads to the scarring patterns that appear in individual wave functions\cite{Heller}. However, usually, scars are weak addition to the random background, and their effect within a single wave function diminishes as the semiclassical limit of short wavelength is approached. On the other hand, being a resonance, the contribution of a given scar extends over a large interval of the energy as illustrated in Fig.~3. Thus
 "mode-locking" of several consequative excited states may result in a rather strong scarring pattern.

To clarify the nature of mode-locking in our system, let us assume that within the frequency range of the excited states there is a single dominant scar, so that the corresponding wave functions may be written in the form \cite{Agam94}
\begin{eqnarray}
u_j({\bf r})= a_j\psi_{scar}({\bf r})+\psi_{rand,j}({\bf r}).  \label{dw}
\end{eqnarray}
Here $\psi_{scar}({\bf r})$ is the same scar contribution shared by all wave functions within the resonance and without loss of generality we shall assume that the weights of this contribution, $a_j$, are real and positive. The second  contribution $\psi_{rand,j}({\bf r})$ is the random component mentioned above, which for different wave functions is, essentially, statistically independent. Consider now a superposition of these wave functions:
 \begin{eqnarray}
u({\bf r})= \sum_j q_j u_j({\bf r}).
\end{eqnarray}
This superposition exhibits mode locking if the weights $q_j$ have the same phase, since in this situation the scar contribution adds up coherently and overwhelms the incoherent contribution of the statistically independent random components.

Now, mode locking will be generated if the resulting excitation pattern is stronger than other possible random like patterns generated by different combinations of the eigenstates $u_j({\bf r})$. The strength of the excitation, in turn, is determined by the interaction matrix elements of the excited state, as follows from formula (\ref{eq:main}). Namely, the smaller the interaction, the stronger the excitation. Thus to close our argument for mode locking, we have to show that the matrix elements of certain scars are smaller than those of random wave patterns.

 Let us consider first the interaction matrix element associated with  a random wave function, denoted henceforth by $I_{rand}$. Approximating such a wave function by a random superposition of plane waves moving in arbitrary directions, we obtain from  formula (\ref{eq:InteractionTerms})
 \begin{eqnarray}
 I_{rand} \sim 3/{\cal A}, \label{hrand}
 \end{eqnarray}
  where ${\cal A}$ is the system's area.

  Consider now the interaction matrix element associated with a stripe pattern (denoted by $I_{st}$) that extends, essentially, all over the system. The wave function associated with this pattern, $u({\bf r}) \sim \sqrt{\frac{2}{\cal A}}\cos({\bf k}\cdot {\bf r})$, may be viewed as two plane waves moving in opposite directions. The resulting matrix element,
  \begin{eqnarray}
 I_{st} \sim 3/(2{\cal A}), \label{hst}
 \end{eqnarray}
is half that of the random pattern. Therefore excitation of stripes, in general, is preferable over the excitation of random patterns. Now, both scars, the  horizontal (Fig.~1a) and the bouncing ball scar (Fig.~1b), being associated with self retracing orbits representing waves moving in opposite directions, result in a local stripe pattern. Therefore whenever the scar contribution, $\psi_{scar}({\bf r})$, is associated with one of these orbits, the system  will be mode-locked in order to generate the stripe pattern.

Yet, this argument is based on the assumption that the stripe pattern of the scars extends over the whole system. In practice the effective area of this pattern, ${\cal A}_{eff}$, is smaller, and therefore  ${\cal A}$ in Eq. (\ref{hst}) should be replaced by the proper effective area of the scar pattern. The bouncing ball scar occupies the central zone of the billiard whose area is approximately ${\cal A}_{eff}=4 {\cal A}/(4+\pi)$. Thus the interaction matrix element of this scar is
 \begin{eqnarray}
 I_{bb} \sim \frac{3(4+\pi)}{8{\cal A}}.
 \end{eqnarray}
This matrix element is still smaller than that the of the random wave function (\ref{hrand}), and therefore the excitation of the bouncing ball stripe pattern is stronger than that of a typical random pattern.
Yet, the bouncing ball scar represents a special case, since it is associated with a continues family of periodic orbits bouncing between the two straight edges of the stadium billiard. In general, the effective area of a scar coming from a generic (isolated) periodic orbit shrinks as the excitation frequency increases. In particular, the effective area occupied by the horizontal scar is of order ${\cal A}_{eff} \sim \sqrt{\lambda L^3}$ where $\lambda$ is the wavelength and $L$ is the length of the billiard. This estimate for the scar's area is explained in Fig.~4.

\begin{figure}[t]
\centering
\includegraphics[width=7.0cm]{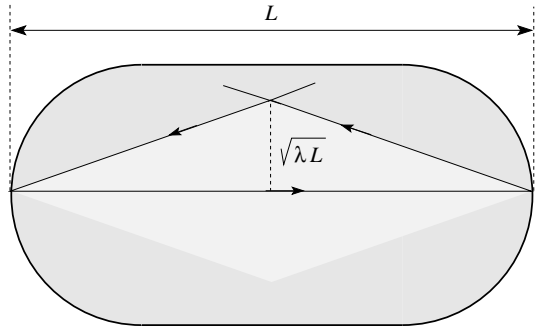}
\caption{The effective area of a scar: The scar wave function is associated with closed orbits that are small deformations of the central periodic orbit. Here we approximate such an orbit as an isosceles triangle whose base is the original horizontal periodic orbit.  The triangle height of the longest orbit that contribute to the scar is determined from the requirement for constructive interference of the orbits, thus the total length of the longest orbit is up to one wavelength $\lambda$ longer than that of the periodic orbit. This conditions implies that that the height is $\sqrt{\lambda L}$ and therefore the effective area of the scar is $L\sqrt{\lambda L}$.}
\label{Fig4}
\end{figure}
Replacing ${\cal A}$ in Eq.~(\ref{hst}) by the effective area we obtain that  the matrix element of the horizontal scar is
\begin{eqnarray}
 I_{h} \sim \frac{3( 4+ \pi)^{3/4}}{16 \sqrt{\lambda {\cal A}^{3/2}}}. \label{hh}
\end{eqnarray}
To obtain this result we have used the relation ${\cal A}= (L^2/16)(4+\pi)$ between the area and the length of the billiard.

Comparing this matrix element with that of the random pattern (\ref{hrand}) it is evident that for sufficiently small $\lambda$, i.e. high enough excitation frequency, $I_{rand}<I_{h}$. Thus the horizontal scar pattern is preferable when $\lambda > (4+ \pi)^{3/2}\sqrt{\cal A}/256\sim L/30$, or at the frequency range where
\begin{eqnarray}
 \omega < \frac{ 2048 \pi c}{ (4+ \pi)^2 L} \simeq 126  \frac{c}{L} \label{range}
\end{eqnarray}
where $c \simeq \sqrt{ \mu/m}$ is the sound velocity.

What about other scars? Our numerical study indicates that scars different from those shown in the upper panels of Fig.~1, are rarely excited.  One of these rare examples is the bow tie scar shown in Fig.~1c. It is somewhat difficult to associate this scarring pattern with the classical orbit shown in the lower right inset, however a Fourier transform of this wave pattern confirms that it is approximately composed of four components of plane waves at directions which differ by $90^0$. Such a combination of plane waves results in square pattern and assuming it to extend over the whole billiard area, we obtain that the corresponding interaction matrix element is $I_{sq}= 9/(4{\cal A})$. It is larger than that associated with stripes but smaller than the matrix element of random wave function. Therefore, in a frequency regime
which does not contain the frequencies of the horizontal and the bouncing-ball scars, such a scarring pattern may emerge. Notice, however, that similar to the horizontal scar, at high excitation frequencies the effective area of this scar diminishes and the random wave pattern becomes preferable.

Longer scars are associated with rays which move in many directions and therefore, their interaction matrix elements become closer to that of the random wave pattern and thus unlikely to be observed. For instance hexagonal pattern emerging from from six plane waves yields an interaction matrix element $I_{hex}= 5/(2{\cal A})$.

To conclude, the horizontal and bouncing ball scars are the dominant excitation patterns because they generate stripes oriented perpendicular to these self retracing orbits.
 Longer periodic orbits generate more complicated density pattern due to interference of waves moving in several directions, e.g.
 at the vicinity of self crossing points of the periodic orbits, or near the boundary of the system where the incidence angle of the impinging orbit differs from zero. Near such points, the local scarring pattern is expected to more complicated. For example in the vicinity of a self crossing point where the orbit crosses itself at right angle the local density excitation will be in the form of a square pattern. Interactions do not favor these patterns and therefore scarring patterns associated with such orbits are less likely to be excited. Finally, all scars associated with isolated periodic orbits disappear at sufficiently high frequencies. The bouncing ball scar is an exception since its effective area is independent of the frequency.

\section{Noise and dephasing}
\label{sec:noise_and_dephasing}

The above conclusions have been drawn under the assumption that the boundary of the billiard plays a crucial role in selecting the patterns of the excited waves in the system. This is because these patterns emerge from the single particle wave functions $u_j({\bf r})$, see Eq.~(\ref{eq:eigenstates}). Yet, when the system is large enough one may expect the boundary to play a minor role, and the local interactions among waves to be the dominant mechanism of pattern selection.

What is the criterion of a large system? Pure dissipation in the manner included in Eq. (\ref{eq:GP}) does not introduce a length scale which competes with the size of the system. Increasing the dissipation mainly lifts the excitation threshold but does not affect the role of the boundary. On the other hand, being an outcome of interference effect, scars are very sensitive to dephasing, namely external, random-like, forces which affect the phase coherence of the system.

Dephasing emerges, for example, in the presence of noise, associated with a stochastic time-dependent component of the confining potential, $\delta V({\bf r},t)$, or interaction with the incoherent component of the condensate \cite{FeshbachNoise}. This introduces an additional length scale to the problem: The dephasing length $L_\phi$ which is the typical distance over which a wave losses its coherence. When the dephasing length is much smaller than the system size, $L_\phi \ll L$ , the BEC density patterns will be governed by local interactions between waves, and the role of the boundary shape will diminish as the system size increases. In this case, as we showed previously, the main candidates for the local patterns are stripes which represent the smallest interaction matrix element.

This implies, in particular, that the horizontal and the bouncing-ball scars of the stadium billiard are rather stable against dephasing, while the bow tie scar shown in Fig.~1c is much less stable. Fig.~1d shows the excitation pattern obtained by adding noise to the system which otherwise generate the pattern shown in Fig.~1c. Fourier transform confirms that this pattern is dominated by stripes. The asymmetry of this pattern is a clear indication that dephasing effects take place. Asymmetric excitation patterns appear also  when adding noise to the horizontal and bouncing ball scars, however, the noise power needed in this case is stronger by two orders of magnitude than that required in order to destroy the pattern shown in Fig.~1c.

\section{Numerical study}
\label{sec:numerical_study}

To check the above conclusions, we have performed an extensive numerical \cite{GPESolve} study of Eq.~ (\ref{eq:GP}). Some of the results of this study are shown in Fig.~1 where we considered a standard stadium billiard ($R=12.0$ microns), with $N=5 \times 10^4$ $^{87}Rb$ atoms, and assuming a strong (trapping frequency of $8.5$ kHz) confinement in the ${\bf z}$ direction. The convergence to steady state excitation is slow, and a rather long ($\sim 500/\omega$) transient response of the system to the parametric drive is generic. We find an exponential sensitivity of the excited modes to $\Gamma$ and $a$, with thresholds clearly visible. We also confirm that the horizontal and the bouncing ball scars patterns are the most likely ones to be excited.

Dephasing has been introduced to the system by adding a time dependent contribution to the confining potential which is random both is space and time and short range correlated in both cases. We verify that as the dephasing noise increases, scars become less prominent, and for small size systems, the BEC density patterns do not respect the symmetry of the system and are dominated by stripes, as demonstrated in Fig.~ 1(d). These patterns are stable against changes in the strength of the noise, and its correlation length, as long as the dephasing length is smaller than the size of the system.

As an additional check of the implication of our theory, we study a different geometry characterized by the symmetry of the dihedral point group $D_4$. The boundary of this chaotic system, formed by four circles, is shown in the left panel of Fig. 5, together with a few of its short periodic orbits. The symmetry of the system generates an exact degeneracy of some of its eigenstates. For instance, a degeneracy between pairs of states scarred along the two shortest periodic orbits of the system (shown by the solid lines in Fig. 5). This degeneracy implies that both states should be excited simultaneously. However, the two scars cross in the middle of the system and therefore a simultaneous excitation of both of them will result in a square pattern near the central point of the system. On the other hand, our analysis of the interaction matrix elements suggests that stripe patterns, associated with the excitation of a single scar, are preferable over the square pattern resulting form the excitation of both scars. Thus one expects to observe excitation patterns in which spontaneous breakdown of symmetry occurs and one scar is selected. This is indeed the case in many patterns parametrically excited in the system. An example is shown in the right panel of Fig.~5.

\begin{figure}[t]
\centering
\begin{tabular}{cc}
      \resizebox{60mm}{!}{\includegraphics[width=7.0cm]{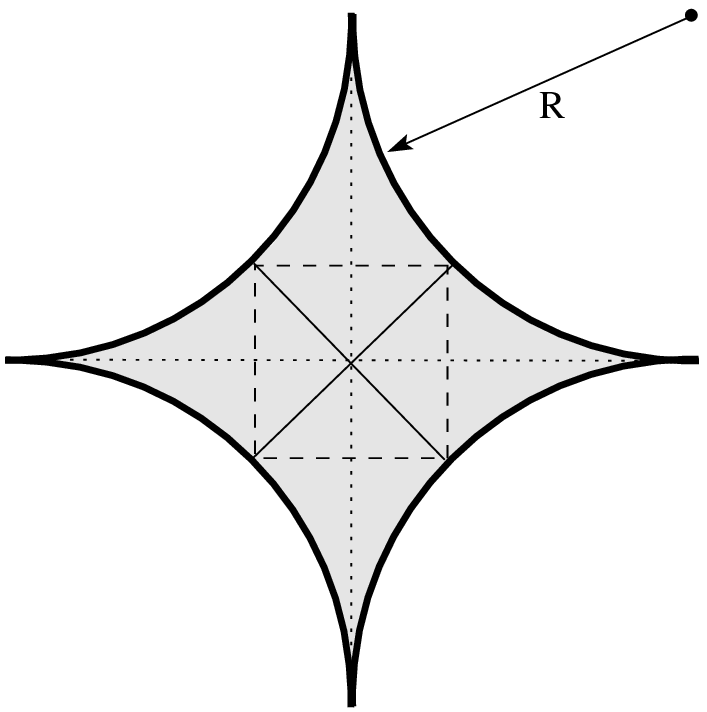}} &
      \resizebox{60mm}{!}{\includegraphics[width=7.0cm]{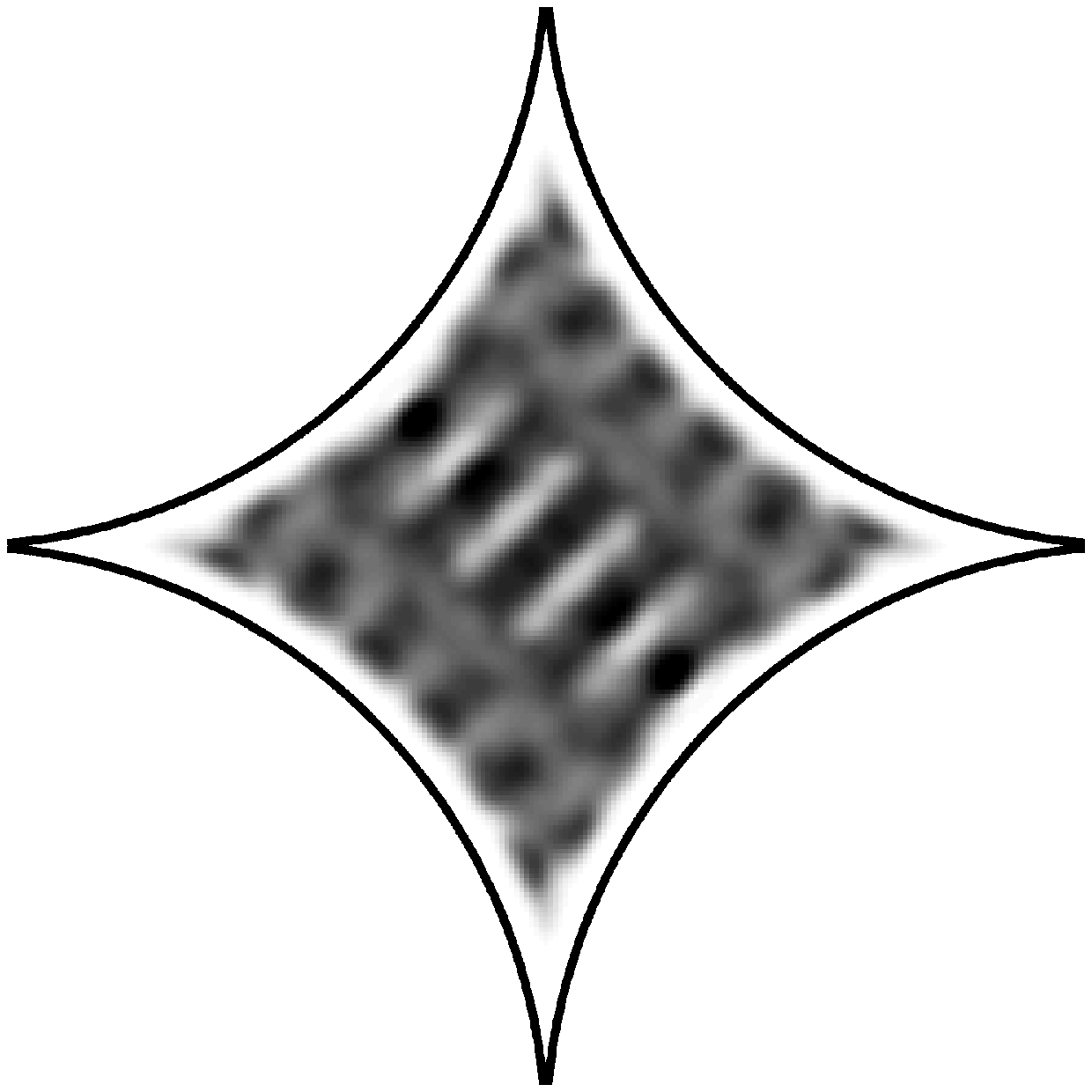}} \\
    \end{tabular}

\caption{Left panel: A billiard made of four touching circles, and the corresponding short periodic orbits of the system. Right panel: A typical density pattern (same number of atoms and density as in Fig. 1, $\omega=2.86 \mu/\hbar$, $a = 0.14 \bar{g}$ and $\Gamma = 0.03$, reflecting the spontaneous breakdown of symmetry between the scars associated with the shortest periodic orbits of the system (the solid lines in the left panel).}
\label{Fig5}
\end{figure}

Yet, the interaction matrix element is not the sole factor which determines the form of the excitation pattern. The precise value of the excitation frequency, for instance, also plays an important role, as follows from formula (\ref{eq:main}). Accordingly, in some of the cases we observe other patterns associated with the short periodic orbits of the system, as shown in the upper row of Fig.~6. Our numerical simulations cannot answer the question whether these are metastable patterns or the the final excitation patterns. But, when introducing noise into the system, these patterns become unstable and taken over by the stripe pattern associated with one of the shortest periodic orbits, see lower row of Fig.~6.

\begin{figure}[t]
\centering
\includegraphics[width=8.0cm]{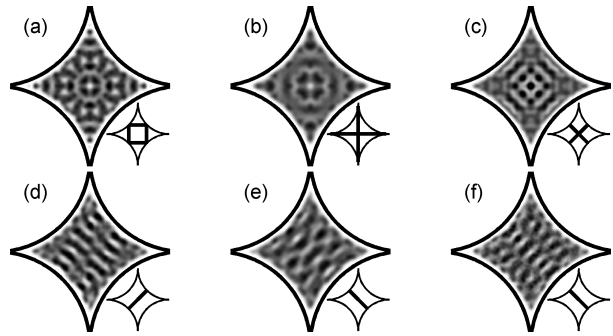}
\caption{Parametric excitation patterns of BEC in the potential well shown at the left panel of Fig.~5.  In the upper row of the figure depicted density plots of the excited patterns at three different excitation frequencies (from left to right, $\omega= 3.42, 2.64, 3.7 \mu/\hbar$) but the same damping and amplitude of excitation ($a = 0.14 \bar{g}$ and $\Gamma = 0.03$). Below each one of these panels we show the density plot obtained for the same parameters but in the presence of noise which generates dephasing.}
\label{Fig6}
\end{figure}

\begin{figure}[h]
\centering
\includegraphics[width=7.0cm]{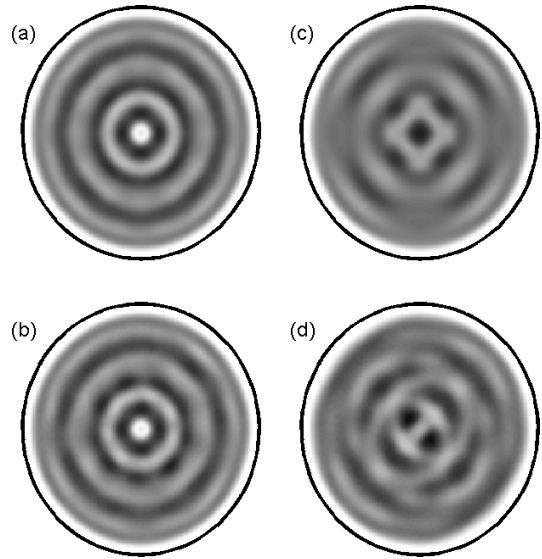}
\caption{Parametric excitation patterns for a circular billiard. We use the same parameters as Fig. 6, with $\omega= 2.7 \mu/\hbar$ in (a) and (c) and $\omega= 3.2 \mu/\hbar$ in (b) and (d). In (c) and (d) we also apply noise as in Fig. 6}
\label{Fig7}
\end{figure}

Finally, it is instructive to compare these results to the case of billiard in which the underlying dynamics is integrable, such as a circular billiard. In Fig. 7 we show the results of such a simulation at two different drive frequencies (a-b), including also the effects of noise (c-d). We observe that the structure of the parametric excitation is not sensitive to noise (up until it is completely destroyed by it). This is consistent with our view that local stripe structure is very stable. In this billiard the stripes only significantly interact at the origin. Therefore noise only causes local changes near the origin (see Fig. 7b and 7d).

\section{Summery}
\label{sec:summery}

To conclude, we show that scars in BECs result from mode-locking of many excited eigenstates. This effect is particularly efficient for the very short scars which exhibit stripe patterns. Longer scars, associated with waves moving in many directions, are hindered by nonlinear effects.  Except for cases like the bouncing ball scar, or more generally scars associated with continuous families of periodic orbits, all other scars are expected to disappear at sufficiently high frequency (small wavelength). This implies that random wave patterns in BEC's may appear, in principle, only at sufficiently high excitation frequencies. Noise also hinders the formation of patterns other than stripes. The latter property may be exploited in order to measure the dephasing time in BECs.

\section*{Acknowledgments}

This work was supported by the United States-Israel Binational
Science Foundation (BSF) grant no.~2008278, and the Israeli Science Foundation (ISF) grant no.~1835/07.



\begin{thebibliography}{99}
\bibitem{Parameters} The parameters used in this figure are:
The number of atoms in the system is $5 \cdot 10^{4}$. The radius of the semicirculs
of the stadium billiard is $R= 12.0 \sqrt{2} \xi$ where $\xi=\hbar/\sqrt{ m \mu }$ is the healing length, and the distancs between the centers of the two semicircles is $2R$.  In panel (a) the
driving frequency of the parametric excitaton is $\omega = 2.6\mu/\hbar$, the drive strength is  $a = 0.08 \bar{g}$, and the dimensionless dissipation parameter is $\Gamma = 0.02$. In panel (b)  parameters same as (a) except for $\omega = 2.85\mu/\hbar$. In panel (c) $\omega=3.218 \mu/\hbar$, $a = 0.07 \bar{g}$ and $\Gamma = 0.015$.
\bibitem{FeshbachNoise} Noise may also appear in the interaction term due to imperfections in the control magnetic field (in the case of a Feshbach resonance). However, this noise is normally not spatially varying. Therefore, its contribution to dephasing of our spatially extended states is negligible.
\bibitem{GPESolve} We solve on a two dimensional 256X256 grid using split Fourier operator propagation as in \cite {Bao}. The ground state is found by imaginary time evolution.
\end{thebibliography}
\end{document}